\documentclass[pra,twocolumn,showpacs,floatfix]{revtex4}
\usepackage{graphicx}
\begin{document}

\title {Manipulating Bose-Einstein condensed atoms in toroidal traps}
\author{A. D. Jackson$^1$ and G. M. Kavoulakis$^2$}
\affiliation{$^1$Niels Bohr Institute, Blegdamsvej 17, DK-2100,
Copenhagen \O, Denmark \\
$^2$Mathematical Physics, Lund Institute of Technology, P.O.
Box 118, SE-22100 Lund, Sweden}
\date{\today}

\begin{abstract}

We consider Bose-Einstein condensed atoms confined in a
toroidal trap. We demonstrate that under conditions of
one-dimensional behavior, the density distribution of the atoms
may be exponentially localized/delocalized, even for very small
variations in the trapping potential along the torus. This
observation allows one to control the atom density externally
via slight modifications of the trapping potential. For similar
reasons, small irregularities of the trap may also have a very
pronounced effect on the density of the cloud.

\end{abstract}
\maketitle

\section{Introduction}

One of the many interesting consequences of recent experimental
advances in the field of cold atoms is the fact that it is now
possible to realize states of reduced dimensionality.  More
precisely, so long as the energy of the atoms due to their
interactions is much smaller than the spacing of the energy
levels associated with motion along some symmetry axis of the
trap, the atomic state is dominated by the lowest state of the
trapping potential, and this degree of freedom is effectively
frozen. Thus, oblate/elongated traps can be used to achieve
quasi-two/quasi-one dimensional behavior.  As reported in
Ref.\,\cite{MIT}, Ketterle and co-workers have achieved
quasi-one-dimensional conditions.  Using a variety of
sophisticated techniques, experimentalists have produced traps
spanning a wide range of other geometries, as reported in,
e.g., Refs.\,\cite{Lag,Dal,Sa,Cr,DSK,Arn}.  Several
experimental and theoretical groups have also studied
quasi-one-dimensional Bose-Einstein condensates in magnetic
waveguides \cite{Mtr}. In a detailed study, Leboeuf and Pavloff
have considered the effects of perturbing obstacles on the
propagation of Bose-Einstein condensed atoms through such
magnetic waveguides \cite{LP}.

In the present study we focus on tight toroidal trapping
potentials, which have been realized in various laboratories
\cite{Sa,Cr,DSK,Arn}. Such traps can be used to create
one-dimensional systems satisfying periodic boundary
conditions. Since such systems have been under investigation
for decades, it is clear that they are interesting from a
theoretical point of view. Further, they may have numerous
technological applications.

In a one-dimensional toroidal trap, the atomic density is
homogeneous if the effective interaction between the atoms is
repulsive \cite{TG}.  It has been shown, however, that for a
sufficiently strong attractive interaction, the atoms undergo a
second-order quantum phase transition and form a localized
density distribution \cite{Ueda,GMK}.  (This stands in contrast
to a three-dimensional gas, which simply collapses.)  In the
inhomogeneous phase the atoms benefit from the formation of a
localized ``blob'', which lowers their interaction energy.

The issue to be considered here is the effect of a controlled
change in the trapping potential $V(\theta)$ along the torus as
a consequence of small transverse irregularities. We will see
that even small irregularities can materially alter this simple
picture of localization.  The effects to be considered were
first encountered in early studies of radar.  Hot spots were
observed wherever waveguides were bent.  It was soon recognized
that any such non-uniformity in a quasi one-dimensional
waveguide necessarily leads to exponentially localized,
subthreshold resonances (i.e., bound states) with a
correspondingly high local energy density.

In Sec.\,II we describe the problem to be considered in greater
detail, and we present our model in Sec.\,III.  Section IV
contains our results for the ground state and the excitation
spectrum of the gas.  Section V contains a discussion of our
results and presents the conclusions of this study.

\section{Effect of a distorted toroidal potential}

One might imagine that small irregularities in the
one-dimensional potential, $V(\theta)$, do not have any
substantial effect.  The situation is very different in the
quasi one-dimensional systems considered here. Consider first a
free particle moving in an infinitely long waveguide of small
transverse size.  The wave function for this particle can be
approximated as the product of the lowest transverse
eigenfunction and a function of the distance along the
waveguide. The associated eigenvalue of the transverse
Hamiltonian provides an effective potential for the resulting
one-dimensional motion.  A localized broadening (or,
equivalently, a bend) of an otherwise uniform waveguide will
thus result in an attractive effective potential.  It is
familiar from elementary quantum mechanics that every purely
attractive potential in one dimension has at least one
exponentially bound state.  The ubiquity of bound states in
infinite waveguides has been investigated thoroughly in an
elegant paper by Goldstone and Jaffe \cite{Jaffe}.

The situation is extremely similar for the toroidal geometry
considered here.  Construct $V(\theta)$ using the same ansatz
for the wave function, and set the zero of the energy as the
maximum value of $V(\theta)$. Assuming a constant longitudinal
wave function, an elementary variational calculation
immediately shows that the ground state energy of this system
is necessarily negative if $V(\theta)$ is non-uniform.  There
will always be classically allowed and forbidden regions with
the usual attendant exponential enhancement or suppression of
the ground state wave function.

For zero or weak interactions, the order parameter for the
atomic system is simply the eigenfunction of the lowest energy
state, and it is exponentially localized/delocalized, too.
Thus, even small irregularities in $V(\theta)$ in such
one-dimensional systems render instability to an inhomogeneous
state inevitable, provided only that the interaction is not too
strong.  Contrary to the quantum phase transition induced by an
attractive effective interaction in a uniform toroidal trap,
irregularities require a sufficiently strong and repulsive
interaction for the density to be (nearly) homogeneous.
Furthermore, while the density variation induced by an
attractive interaction in the uniform case is sinusoidal close
to the transition point, irregularities in $V(\theta)$ result
in an exponential localization/delocalization, which is
generally more pronounced.

\section{Model}

To demonstrate these effects quantitatively, we consider the
following form of $V(\theta)$:
\begin{eqnarray}
  V(\theta) = \left\{ \begin{array}{ll}
  V_0 & \rm{ for \,\, }  |\theta| > \pi/10 \\
  0 & \rm{ for \,\, } |\theta| < \pi/10,
  \end{array} \right.
\label{potential}
\end{eqnarray}
which corresponds to a broadening of the torus over 10\% of its
circumference.  We assume the usual contact potential for the
atom-atom interaction, $V_{\rm int}({\bf r}-{\bf r}') = U_0
\delta({\bf r}-{\bf r}')$ with $U_0 = 4 \pi \hbar^2 a /M$.
Here, $a$ is the scattering length for elastic atom-atom
collisions and $M$ is the atomic mass. The Hamiltonian of the
quasi-one-dimensional system thus becomes
\begin{eqnarray}
  H/N = &-& \frac {\hbar^2} {2MR^2} \int \Psi^{\dagger}(\theta)
  \frac {\partial^2} {\partial \theta^2} \Psi(\theta) \, d \theta +
  \nonumber \\
  &+& \int \Psi^{\dagger}(\theta) V(\theta) \Psi(\theta) \,
  d\theta +
  \nonumber \\
  &+& \frac {2 \pi \hbar^2 N a} {M R S} \int \Psi^{\dagger}(\theta)
  \Psi^{\dagger}(\theta) \Psi(\theta) \Psi(\theta) \, d \theta.
\label{Ham1}
\end{eqnarray}
Here $N \gg 1$ is the number of atoms.  The prefactor of the
last term, $2 \pi \hbar^2 N a/(M R S)$, is equal to $\pi n
U_0$, where $n$ is the (three-dimensional) density, $n=N/(2 \pi
R S)$, $R$ is the radius of the torus, and $S$ its cross
section, with $\sqrt S \ll R$. The three energy scales that
appear in our problem, are then (i) the energy for motion along
the torus $\hbar^2/ (2 M R^2)$, (ii) the depth of the potential
$V_0$, and (iii) the interaction energy between the atoms, $n
U_0$.

For $V_0 = 0$, one finds an instability towards the formation
of a localized state induced by an attractive interaction when
$n |U_0| \sim \hbar^2/(2 M R^2)$. More precisely, the critical
value of the ratio, $\gamma$, between the potential energy
$nU_0$ and the kinetic energy $\hbar^2/(2MR^2)$, is $\gamma = 4
N a R/S = -1/2$ \cite{Ueda,GMK}. Finally, the crossover between
the phases of localized and homogeneous density that we
consider here is given roughly by the condition $n U_0 \sim
V_0$.

\begin{figure}[t]
\includegraphics[width=7cm,height=4.2cm]{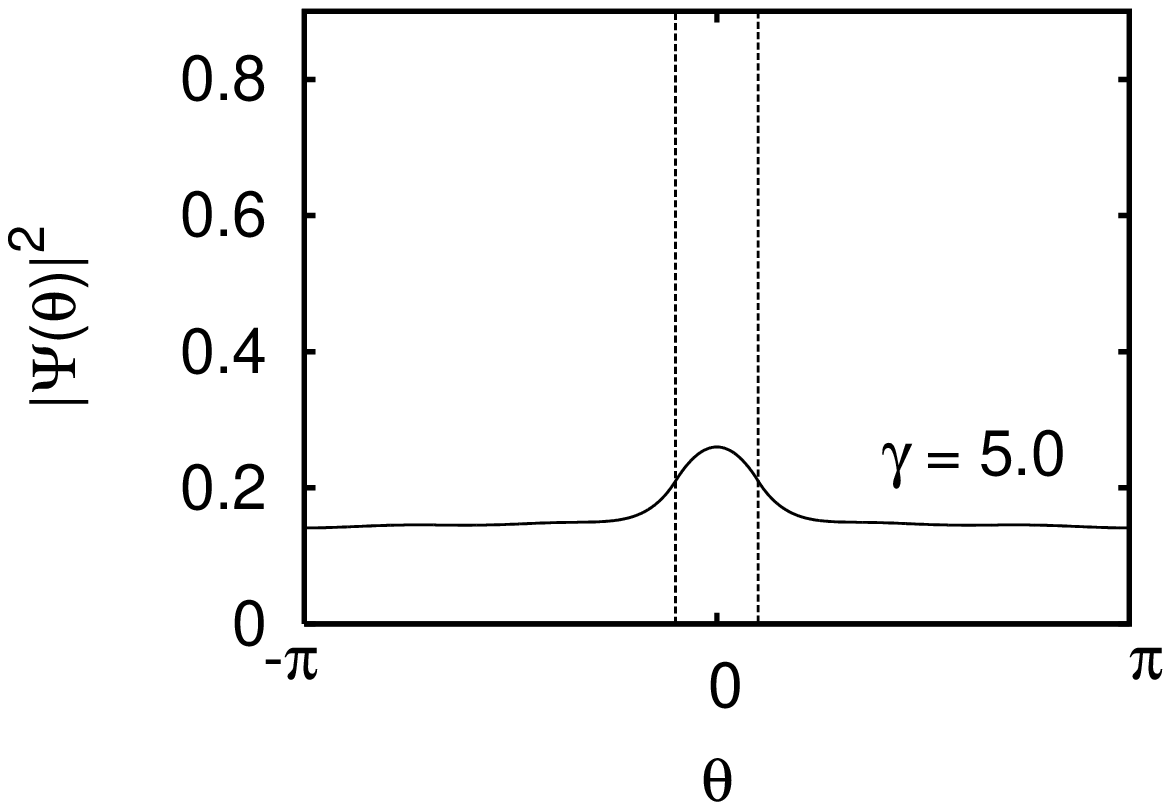}
\includegraphics[width=7cm,height=4.2cm]{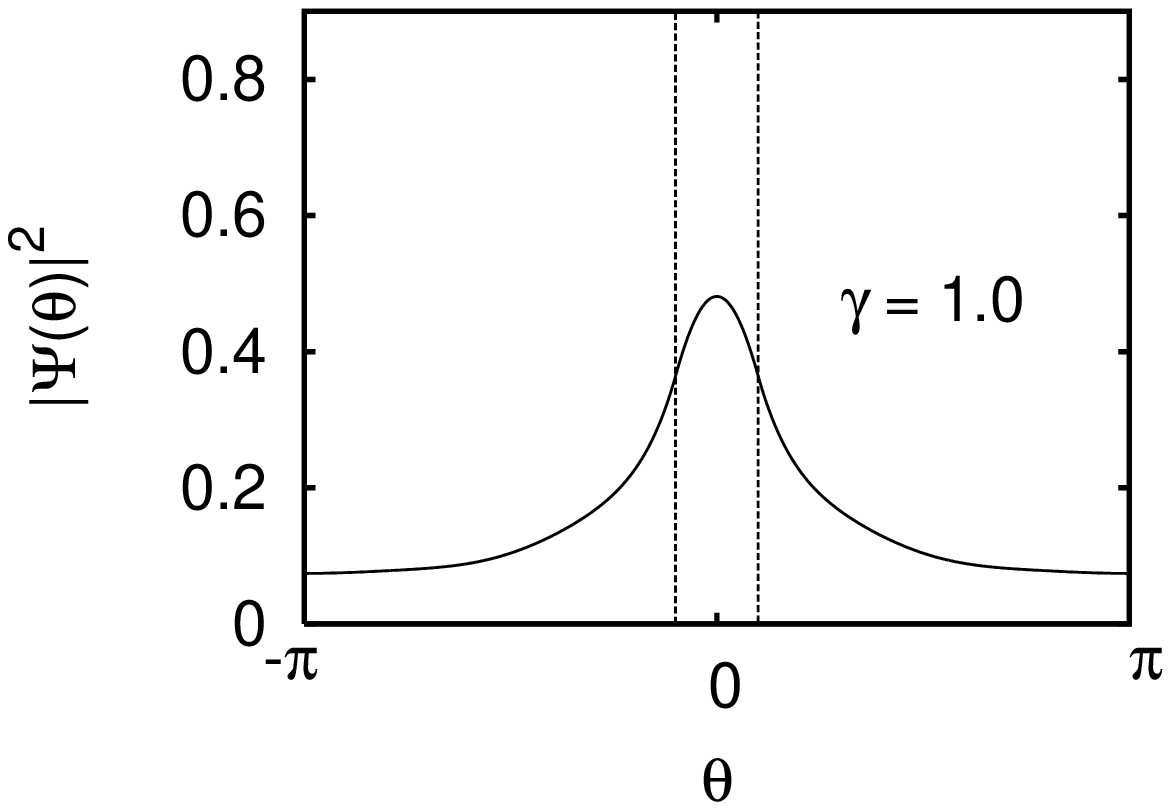}
\includegraphics[width=7cm,height=4.2cm]{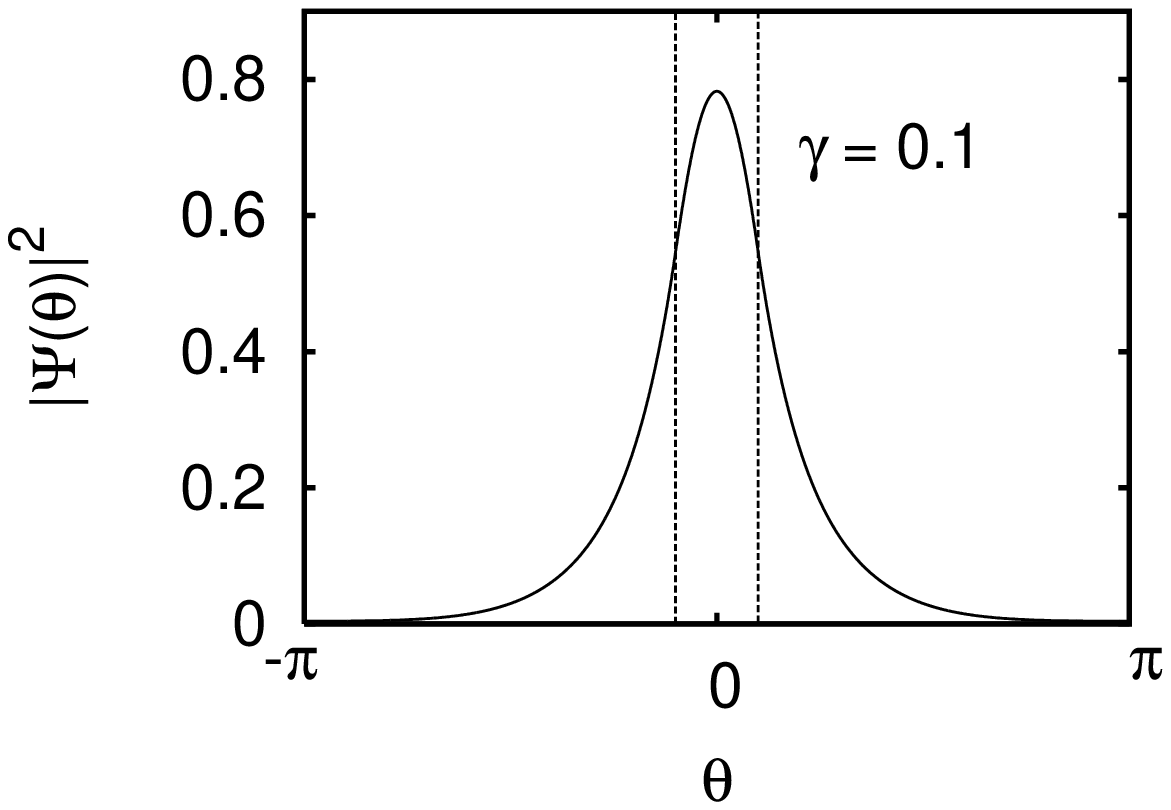}
\caption[]{The density $|\Psi(\theta)|^2$ of Eq.\,(\ref{op})
for three values of the coupling constant $\gamma = 0.1, 1.0$,
and 5.0, and for $V(\theta)$ given by Eq.\,(\ref{potential}).
The two vertical lines show the range of the potential
$V(\theta)$.} 
\label{FIG1}
\end{figure}

\section{Results}
\subsection{Lowest state}

To determine the order parameter $\Psi(\theta)$, we first solve
the eigenvalue problem $H_0 \psi_m = E_m \psi_m$, where $H_0$
is the Hamiltonian of the non-interacting problem. The
eigenstates $\psi_m$ are parity eigenstates. The actual value
of $V_0$ that we choose is $50/\pi^2$ in units of
$\hbar^2/(2MR^2)$, and it has only one bound state with an
energy $E_0 \approx 3.503 \hbar^2/(2MR^2)$. Except for this
lowest ($m=0$) eigenvalue, the eigenvalues of all the excited
states are quite close to those of the undistorted torus, i.e.,
\begin{equation}
  E_m \approx V_0 + \frac {\hbar^2} {2 M R^2} \, m^2.
\label{ene}
\end{equation}
Clearly, this expression becomes more accurate for larger
values of $m$.

Having solved the eigenvalue problem, we then expand the order
parameter in the eigenstates $\psi_m$,
\begin{eqnarray}
  \Psi (\theta) = \sum_m c_m \psi_m(\theta),
\label{op}
\end{eqnarray}
and calculate the coefficients $c_m$ variationally.  For each
value of the coupling constant $\gamma$ considered, we minimize
the expectation value of the Hamiltonian subject to the constraint
$\sum_m |c_m|^2 = 1$ corresponding to a fixed number of atoms.

Figure 1 shows $|\Psi(\theta)|^2$ for three values of $\gamma =
0.1, 1.0$, and 5.0, where we kept the lowest five states (of
even parity -- only even parity states contribute to $\Psi$).
The highest eigenstate, with quantum number $m_0$, that must be
included in the expansion of the order parameter has to satisfy
the condition $E_{m_0} \gg n U_0$. This condition is
comfortably met even for the largest interaction strength
$\gamma = 5.0$ considered, since $E_5 \approx 29.620
\hbar^2/(2MR^2)$.

For small values of the interaction energy, $n U_0 \ll E_0$,
the dominant component of the order parameter is the lowest
eigenstate of the single-particle problem. This is an
exponentially decaying state in the classically forbidden
regions of the torus, $|\theta| > \pi/10$.  Clearly, a similar
result will be obtained for any non-uniform choice of
$V(\theta)$. As the coupling increases, the density
distribution becomes wider. For sufficiently strong
interactions, $n U_0 \gg V_0$, the density $|\Psi(\theta)|^2$
becomes homogeneous, reaching the limiting value $1/(2\pi)$.

\begin{figure}[t]
\includegraphics[width=7cm,height=4.8cm]{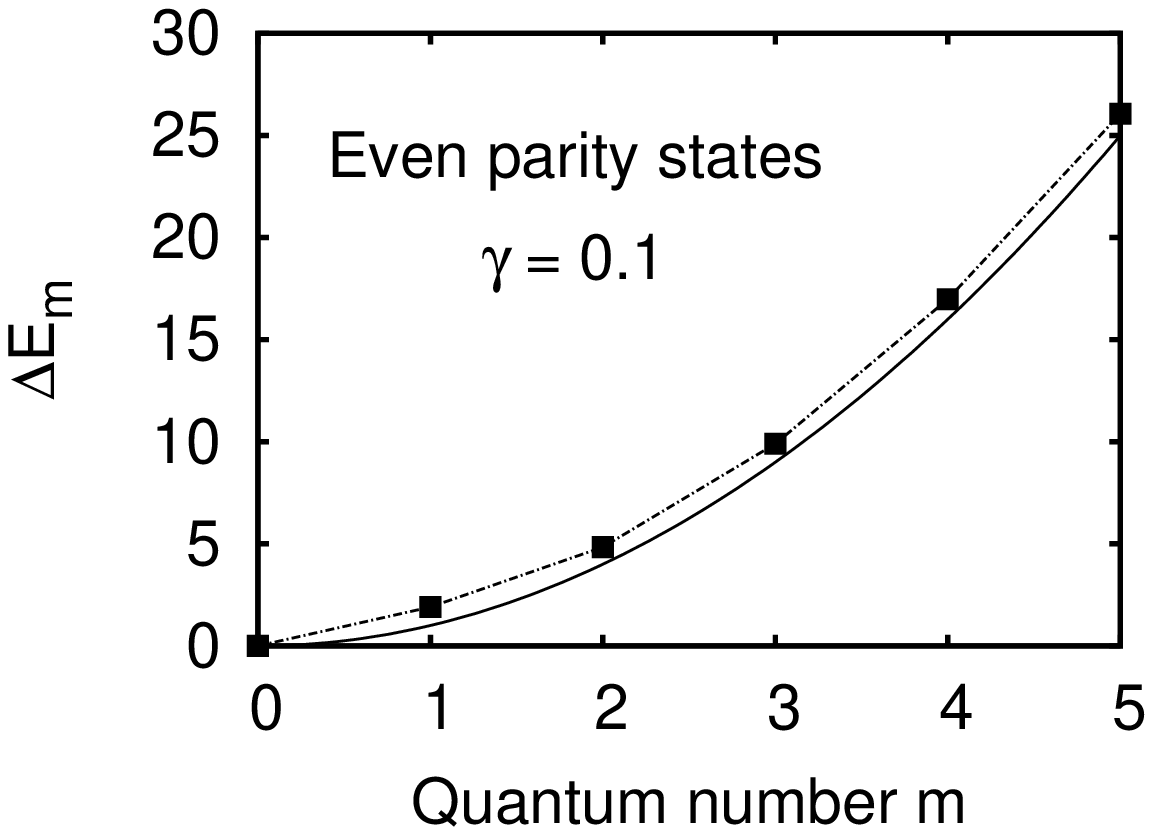}
\includegraphics[width=7cm,height=4.8cm]{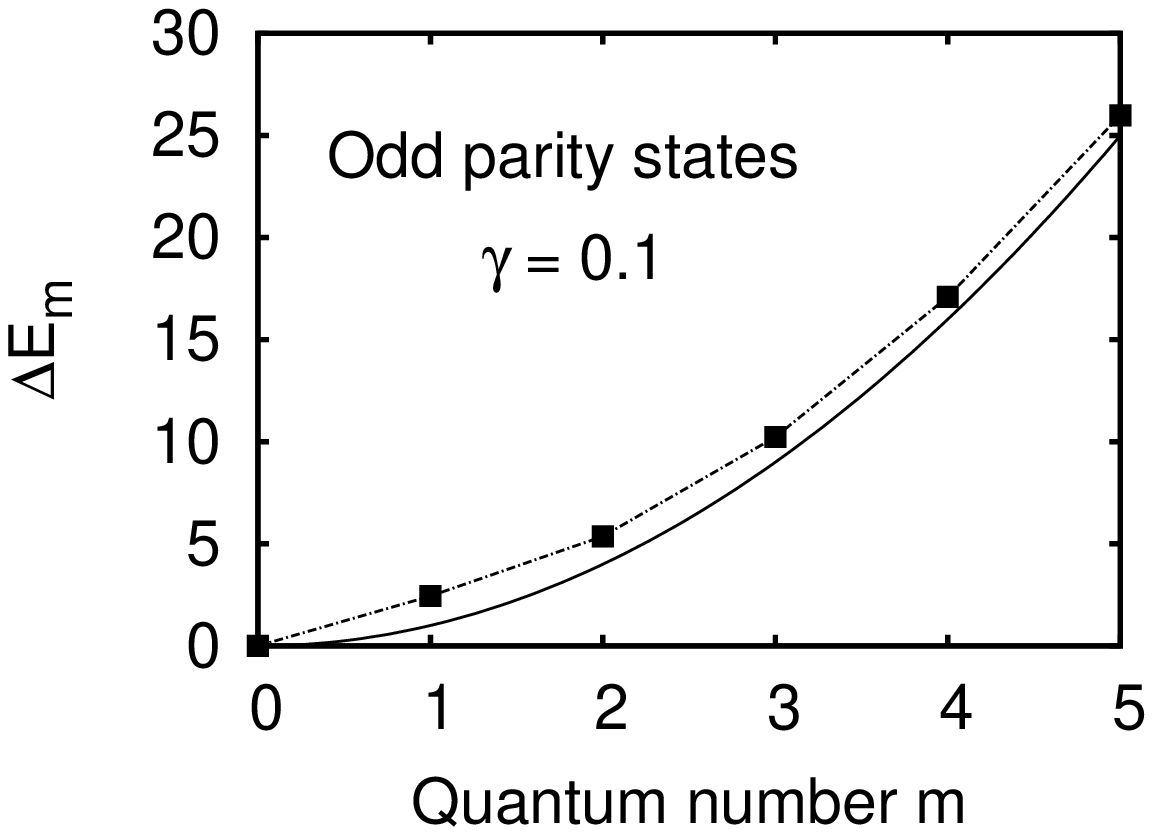}
\caption[]{The points (that are connected with straight, dashed
lines) show the excitation spectrum, Eq.\,(\ref{exc}), for
$\gamma = 0.1$ for the lowest five excited states of even and
odd parity. The solid lines shows the free-particle excitation
energy $E_m - E_0$. The energy is measured in units of
$\hbar^2/(2MR^2)$.}
\label{FIG2}
\end{figure}

\subsection{Excitation spectrum}

The excitation spectrum is also affected as one goes from the
limit of zero interactions (with the order parameter being a
localized state) to very strong interactions (with the order
parameter being a homogeneous state). For weak interactions,
the excitations are single-particle like. Atoms are excited
from the eigenstate of the non-interacting problem with quantum
number $m=0$ to some excited state with quantum number $m \neq
0$. In the absence of interactions the corresponding excitation
energies are approximately equal to the quadratic law of
Eq.\,(\ref{ene}).

For weak interactions ($\gamma \ll 1$) the excitation spectrum
can be calculated using perturbation theory. The total energy
of the lowest state is
\begin{eqnarray}
   {\cal E}_0 &=& N E_0 + \frac {2 \pi \hbar^2 a} {M R S} N (N-1)
   \, I_{0000},
\label{e01}
\end{eqnarray}
where $I_{ijkl}= \int \psi_i^* \psi_j^* \psi_k \psi_l \, d
\theta$. Exciting one atom to the state with quantum number
$m$, the total energy becomes
\begin{eqnarray}
   {\cal E}_m &=& (N-1) E_0 + E_m +
   \nonumber \\
   &+& \frac {2 \pi \hbar^2 a} {M R S}
   [(N-1) (N-2) \, I_{0000} + 4 (N-1) \, I_{0m0m}].
\nonumber \\
\label{e02}
\end{eqnarray}
The factor of four in the last term comes from the direct and
the exchange terms of the interaction energy for identical
bosons. From Eqs.\,(\ref{e01}) and (\ref{e02}) we get for
$\Delta {\cal E}_m = {\cal E}_m - {\cal E}_0$ that
\begin{eqnarray}
   \Delta {\cal E}_m = E_m - E_0 +
    2 \pi \gamma \, \frac {\hbar^2} {2 M R^2}
   [2 \, I_{0m0m} - I_{0000}].
   \nonumber \\
\label{exc}
\end{eqnarray}
Figure 2 shows $\Delta {\cal E}_m$ for the five lowest excited
eigenstates of the non-interacting system of even and odd
parity, for $\gamma = 0.1$. The excitation energies of the
first two excited states (i.e., the even and odd parity states
with $|m|=1$) are $\approx 1.907 \hbar^2/(2MR^2)$ and $\approx
2.443 \hbar^2/(2MR^2)$, respectively.  Clearly for large enough
$|m|$ the excitation spectrum is quadratic. On the other hand,
for the low-lying excited states there are deviations from the
quadratic dependence due to the interactions, and also due to
the deviations of the eigenenergies from the formula of
Eq.\,(\ref{ene}), which are more pronounced for low $m$.

The opposite limit of strong interactions, $\gamma \gg 1$, is
more straightforward. Even in the presence of some $V(\theta)$
the density tends towards homogeneity in this limit (as shown
in the top panel of Fig.\,1). The excitation spectrum of the
homogeneous state, $|\Psi(\theta)|^2 = 1/(2 \pi)$, is
\cite{Ueda,GMK}
\begin{equation}
   \Delta {\cal E}_m = \frac {\hbar^2} {2MR^2}
   \sqrt{m^2 (m^2 + 2 \gamma)}.
\end{equation}
For $\gamma \gg 1$ the above formula implies that $\Delta {\cal
E}_m$ scales as $|m|$ in our problem,
\begin{equation}
   \Delta {\cal E}_m = \frac {\hbar^2} {2MR^2} \sqrt{2\gamma}
   \, |m|.
\label{dr}
\end{equation}
The above equation implies a speed of sound $c = (R / \hbar) \,
\partial \Delta {\cal E}_m/\partial m = \hbar \sqrt{2 \gamma}/
(2MR)$, or $M c^2 = n U_0$.  This result can also be obtained
from the energy per unit length, $\epsilon(\sigma)$, according
to the formula $M c^2 = \sigma \, \partial^2 \epsilon /
\partial \sigma^2$, where $\sigma = N/(2 \pi R)$ is the atom
density per unit length. In the limit of strong interactions,
where $|\Psi|^2 = 1/(2\pi)$, the quadratic contribution to
$\epsilon(\sigma)$ is given by the interaction energy, i.e.,
the last term in the Hamiltonian $H$ of Eq.\,(\ref{Ham1}),
$\epsilon (\sigma) = 2 \pi \hbar^2 a \sigma^2/(M S)$. Thus, $c
= \hbar \sqrt{2 \gamma}/(2MR)$, in agreement with
Eq.\,(\ref{dr}).

\section{Discussion and conclusions}

Depending on the various parameters and on the experimental
demands in constructing axially-symmetric toroidal traps, it is
possible to explore relevant parts of the phase diagram of
quasi one-dimensional atomic gases.  We have pointed out,
however, that the ground state of an atom in a toroidal trap
will always have some degree of exponential localization
whenever the trapping potential deviates from perfect
uniformity.  Although we have illustrated this fact with one
specific form of $V(\theta)$, the result is general.  Even
apparently small modulations in $V(\theta)$ can give rise to
the localization demonstrated here and have the potential to
obscure the quantum phase transition described above for an
effective attractive interaction between the atoms.
Furthermore, the excitation spectrum is also affected by the
presence of some variation in $V(\theta)$. More specifically,
this is free-particle-like for zero or sufficiently weak
interactions, and it is phonon-like for sufficiently strong
interactions.

Our study leads to three conclusions: (i) In realistic toroidal
traps, even weak irregularities can result in a density
distribution that is exponentially localized.  Care must
therefore be exercised in comparing experimental results with
theoretical predictions assuming a uniform trapping potential.
(ii) The excitation spectrum (and thus the dynamics) of the gas
is also affected by variations in $V(\theta)$, and caution is
again called for. (iii) The controlled modification of the
toroidal potential and/or tuning of the coupling constant of
the atomic interaction permit engineering of the shape of the
atom density in a rather dramatic way. For example, the
creation in the torus of a potential resembling our $V(\theta)$
(e.g., using a laser) can change the density distribution from
almost homogeneous to highly localized. Since such
modifications of toroidal traps can be made with relative ease
and that the coupling between the atoms is easily tunable via
Feshbach resonances, our results may have useful applications.

GMK acknowledges financial support from the European Community
project ULTRA-1D (NMP4-CT-2003-505457).


\begin{thebibliography}{99}
\bibitem{TG} For very strong repulsive interactions one gets to
the so-called Tonks-Girardeau limit, of impenetrable bosons. We
do not consider this limit in the present study.
\end{thebibliography}
\end{document}